# Electrical characterization of structured platinum diselenide devices

Chanyoung Yim,[1] Vikram Passi,[1] Max C. Lemme,[1,2] Georg S. Duesberg,[3,4,5] Cormac Ó Coileáin[3] Emiliano Pallechi,[6] Dalal Fadil,[6] and Niall McEvoy,[3,4,*]

[1]Department of Electrical Engineering and Computer Science, University of Siegen, Hölderlinstraße 3, 57076 Siegen, Germany

[2]Faculty of Electrical Engineering and Information Technology, RWTH Aachen University, Otto-Blumenthal-Str. 25, 52074 Aachen, Germany

[3]Centre for the Research on Adaptive Nanostructures and Nanodevices (CRANN) and Advanced Materials and BioEngineering Research (AMBER), Trinity College Dublin, Dublin 2, Ireland

[4]School of Chemistry, Trinity College Dublin, Dublin 2, Ireland

[5]Institute of Physics, EIT 2, Faculty of Electrical Engineering and Information Technology, Universität der Bundeswehr München, Werner-Heisenberg-Weg 39, 85577 Neubiberg, Germany

[6]Institut d'Electronique, de Microélectronique et de Nanotechnologie (IEMN,) 59652 Villeneuve d'Ascq, France

[*]Corresponding author: nmcevoy@tcd.ie

**Abstract**

Platinum Diselenide (PtSe$_2$) is an exciting new member of the two-dimensional (2D) transition metal dichalcogenide (TMD) family. It has a semimetal to semiconductor transition when approaching monolayer thickness and has already shown significant potential for use in device applications. Notably, PtSe$_2$ can be grown at low temperature making it potentially suitable for industrial usage. Here, we address thickness dependent transport properties and investigate electrical contacts to PtSe$_2$, a crucial and universal element of TMD-based electronic devices. PtSe$_2$ films have been synthesized at various thicknesses and structured to allow contact engineering and the accurate extraction of electrical properties. Contact resistivity and sheet resistance extracted from transmission line method (TLM) measurements are compared for different contact metals and different PtSe$_2$ film thicknesses. Furthermore, the transition from semimetal to semiconductor in PtSe$_2$ has been indirectly verified by electrical characterization in field-effect devices. Finally, the influence of edge contacts at the metal – PtSe$_2$ interface has been studied by nanostructuring the contact area using electron beam lithography. By increasing the edge contact length, the contact resistivity was improved by up to 70 % compared to devices with conventional top contacts. The results presented here represent crucial steps towards realizing high-performance nanoelectronic devices based on group-10 TMDs.

**Keywords**: contact resistance, transition metal dichalcogenides, platinum diselenide, nanoelectronic devices

**Introduction**

Over the last decade, the need for further miniaturization and increased functionality of electronic devices has triggered massive research in two-dimensional (2D) channel materials such as graphene and transition metal dichalcogenides (TMDs).[1–3] Reliable electrical contacts between devices/materials and metal electrodes are crucial, as the contact resistance can strongly influence or dominate the behavior of the entire device. Depending on the contact status at the metal – channel junction, Ohmic or Schottky, the flow of charge carriers across the device can be decisively deteriorated.[4] Thus, the reliable formation of contacts to each new channel material proves challenging and has to be investigated in detail.

Graphene has been considered one of the most promising candidates for future nanoelectronics due to its unique electrical properties.[5] Contact resistivity in graphene devices has been widely studied and a wide range of contact resistances (~$10^2$ to ~$10^3$ Ω·µm) has been reported, depending on the contact metal, surface states and contact geometry.[6–9] Following the impressive advances in graphene, various layered 2D TMDs have been tested in a wide range of device applications, including field effect transistors (FETs),[10,11] photodetectors[12–14] and sensors.[15,16] Up to now, molybdenum and/or tungsten based materials have been the main focus of 2D TMD research, and the majority of studies on electrical contacts to 2D TMDs have concentrated on these materials.[17–19] However, compared to classical silicon-based devices, these TMDs still show relatively inferior performance and less environmental stability in device applications.[20–23] Moreover, the high growth temperature (> 600 °C) associated with TMD synthesis by chemical vapor deposition (CVD), which is typically used for large-scale TMD film synthesis,[24–26] can limit their compatibility with current semiconductor processing.

On the other hand, there are other relatively unexplored members of the TMD family such as group-10 TMDs. Recently, the electronic structure and properties of these materials have been theoretically evaluated and, as a consequence of their promising characteristics, they have been proposed for use in electronic device applications.[27,28] Platinum diselenide ($PtSe_2$) is one such group-10 TMD which is known to be a semimetal in bulk form with zero bandgap.[29] Theoretical calculations suggested a transition from semimetal to semiconductor with reduced $PtSe_2$ thickness[27,30] and it was shown experimentally by Wang *et al.* that monolayer $PtSe_2$ has a bandgap of ~1.2 eV.[31] In addition, we recently reported that layered $PtSe_2$ can be synthesized in a scalable manner at low temperature (400 °C) by thermally assisted conversion (TAC) of pre-deposited Pt layers and utilized as the active material in optoelectronic and gas-sensor devices.[32,33] Such low-temperature synthesis may potentially have a high impact on practical device applications since it allows integration of $PtSe_2$ with standard semiconductor back-end-of-line (BEOL) processing.[34–36] In this regard, it is critical to evaluate electrical properties and contacts to $PtSe_2$ in electronic devices.

Here, $PtSe_2$ channels with controlled dimensions and thicknesses were grown using a TAC method. Electron beam lithography (EBL) was used to fabricate transmission line method (TLM) structures to extract contact resistivity and sheet resistance of the $PtSe_2$ devices. In addition, electrical characterization of $PtSe_2$ FETs was conducted to study the charge-transport characteristics of the $PtSe_2$ devices. Finally, we investigated the effect of "edge" or side contacts on the contact resistivity using well-defined hole-patterns in the contact region of the $PtSe_2$ channel.

**Methods**

P-type silicon (p-Si) wafers (boron, $3 \times 10^{15}$ cm$^{-3}$, <100>) with a thermally grown silicon dioxide (SiO$_2$, thickness: 290 nm) layer were prepared as substrates for the devices. An electron beam lithography (EBL) system (Raith EBPG-5000Plus) was used to pattern the PtSe$_2$ channels and contacts. Initial Pt layers with different thicknesses were sputtered onto the substrates using a Gatan coating system (Gatan 682 PECS) with a deposition rate of < 0.1 nm per second, followed by a lift-off process.

A TAC process was used to synthesize layered PtSe$_2$ thin films, as described in our previous work.[32,33] The sputtered Pt samples and the Se source (Sigma Aldrich) were placed in two separate, independently controlled heating zones of a quartz tube furnace. The primary heating zone where the Pt samples were located was heated to 400 °C and the second heating zone for the Se source was heated to the melting point of Se (~220 °C) under Ar/H$_2$ (9:1) gas flow, leading to the formation of layered PtSe$_2$ thin films. Different metal electrodes of titanium/gold (Ti/Au, 20/150 nm) and nickel/gold (Ni/Au, 20/150 nm) were deposited to form TLM structures using an electron beam evaporation system (Plassys) without doing any additional processing before evaporation, followed by a lift-off process.

Edge-contacted structures between the channel and metal electrodes were realized by patterning arrays of holes (with a hole-size of 200 × 200 nm) using EBL on the contact area of the channels, resulting in a contact area with empty holes on the channel after Pt deposition. The metal electrodes were patterned by EBL after selenization of the Pt. The contact metal was evaporated on to the contact region of the PtSe$_2$ channel with holes, without doing any additional processing at the contact area before evaporation.

Raman spectra were recorded with a Witec Alpha 300 R confocal Raman microscope, using an excitation wavelength of 532 nm and a spectral grating with 1800 lines/mm. Scanning electron microscopy (SEM) images were taken using a JEOL SEM (JSM-IT300) at a high-vacuum mode with an accelerating voltage of 2 kV. Electrical measurements were performed at room temperature under ambient conditions using a Karl Süss probe station connected to a Keithley semiconductor analyzer (SCS4200).

**Results and discussion**

PtSe$_2$ device channels were synthesized by direct selenization of pre-deposited Pt layers with different thicknesses. As reported in our previous studies,[32,33] the PtSe$_2$ synthesized by a TAC process has a polycrystalline structure, which is also observed in SEM images (Figure S1 of the Supplementary Information) of the PtSe$_2$ film surface. According to atomic-force microscopy (AFM) measurements of the thicknesses of Pt and layers before and after selenization, it has been found that the initial Pt thickness expands approximately four times after selenization. Details of the AFM characterization are presented in Figure S2 and S3 of the Supplementary Information. A schematic diagram of the film growth process is presented in Figure 1(a). Firstly, the channel area was defined on the substrates by EBL. After Pt deposition and lift-off, the Pt samples were selenized via a TAC method. Raman spectra of the PtSe$_2$ layers grown from various Pt thicknesses (0.5, 2 and 5 nm) are shown in Figure 1(b). Two prominent peaks at ~177 cm$^{-1}$ and ~210 cm$^{-1}$ can be seen in the spectra. They represent the typical Raman fingerprint of layered PtSe$_2$ with a 1T type crystal structure and are related to the E$_g$ (~177 cm$^{-1}$) and A$_{1g}$ (~210 cm$^{-1}$) Raman active modes, respectively. [32] The E$_g$ mode indicates an in-plane vibrational mode of Se atoms and the A$_{1g}$ mode is an out-of-plane vibration of Se atoms. As the films get thicker, a slight red shift of the $E_g$ mode is observed, alongside an increase in the relative intensity of the $A_{1g}$ mode, consistent with previous reports.[32,33] Such an increase in the relative intensity of the $A_{1g}$ mode implies a greater out-of-plane contribution, which may be due to enhanced van der Waals interactions in the thicker films.

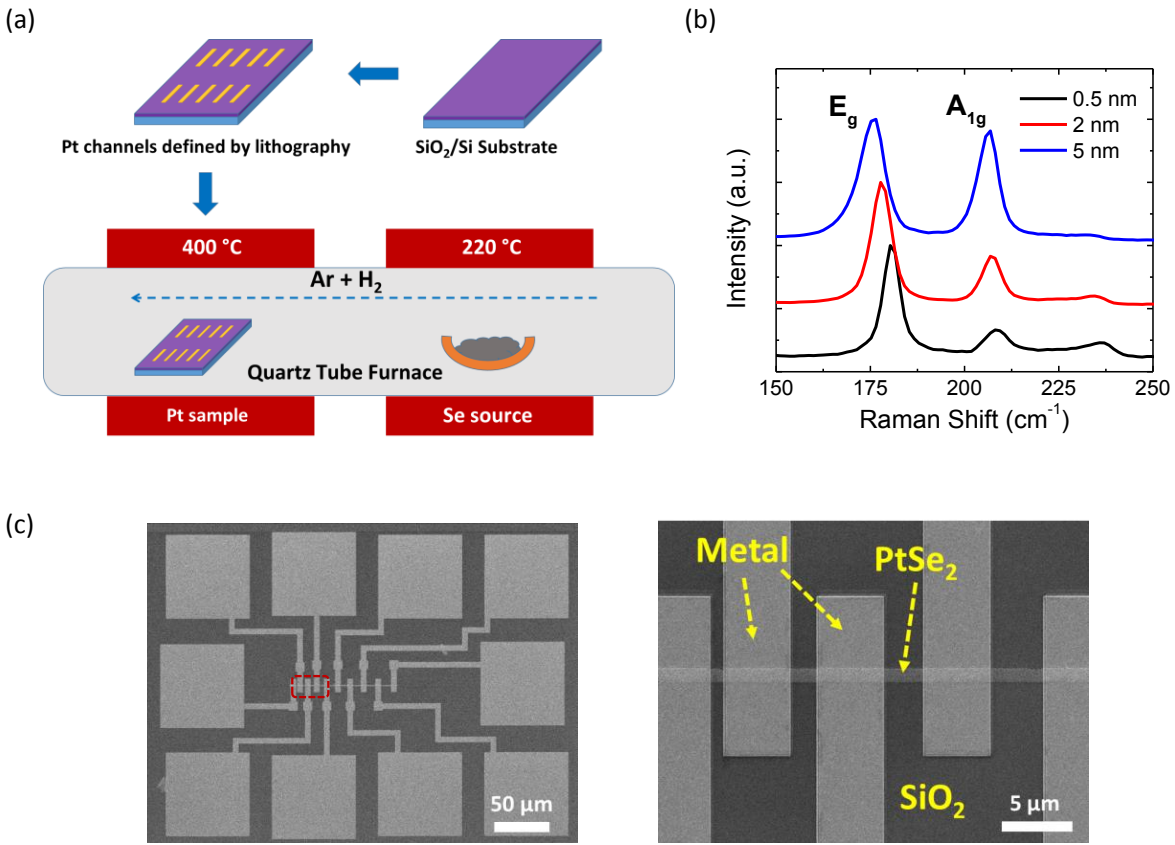

Figure 1. (a) Schematic diagram of the $PtSe_2$ channel synthesis process using a TAC method. (b) Raman spectra of $PtSe_2$ films of different Pt deposition thickness normalized to the $E_g$ mode intensity. Initial Pt deposition thicknesses are 0.5, 2 and 5 nm. (c) SEM image of the fabricated $PtSe_2$ channel device with a TLM structure (left) and its enlarged image (right). The contact spacing increases from 1 to 9 μm with a step of 1 μm.

TLM measurements were used at room temperature to determine the contact resistance of the PtSe$_2$ devices in this work.[37] TLM structures were patterned on the substrates with predefined PtSe$_2$ channels by EBL. The contact spacing was varied from 1 to 9 μm in 1 μm steps. Scanning electron microscopy (SEM) images of a TLM structure on a PtSe$_2$ channel are shown in Figure 1(c). Two representative metals with low and high work function, Ti and Ni, were chosen and used to contact the PtSe$_2$ channels, both of which were selenized from 5 nm thick initial Pt layers. The contact resistances of these were compared through dc current – voltage (I – V) measurements of the TLM structures. Figure 2(a) and (b) show I – V data measured from Ti and Ni contacted PtSe$_2$ TLM structures with a channel width of 3 μm, respectively. The linear I – V curves indicate that both Ti and Ni electrodes form good ohmic contacts with PtSe$_2$ channels. Values of the contact resistance for each metal and the sheet resistance of the PtSe$_2$ channel can be extracted by extrapolation from linear fits of the plots of the total resistance ($R_T$) versus (vs.) the contact spacing (Figure 2(c) and (d)), wherein the y-intercept and the slope of the fits provide information on the contact resistance and sheet resistance. Figure 2(e) shows contact resistivity values of each metal for different PtSe$_2$ channel widths, where the contact resistance values were normalized to channel width for direct comparison. It was found that the Ti-contacted device had a contact resistivity more than one order of magnitude higher than the Ni-contacted device. Considering the work function values of the contact metals (Ti: 4.3 eV, Ni: 5.2 eV), we observe that the metal with a higher work function has a lower contact resistance with PtSe$_2$. The sheet resistance values of the PtSe$_2$ channels in Figure 2(f) are quite similar to each other (500 – 600 Ω/□) for both contact metals, which can be expected from the identical PtSe$_2$ channel thicknesses for both devices.

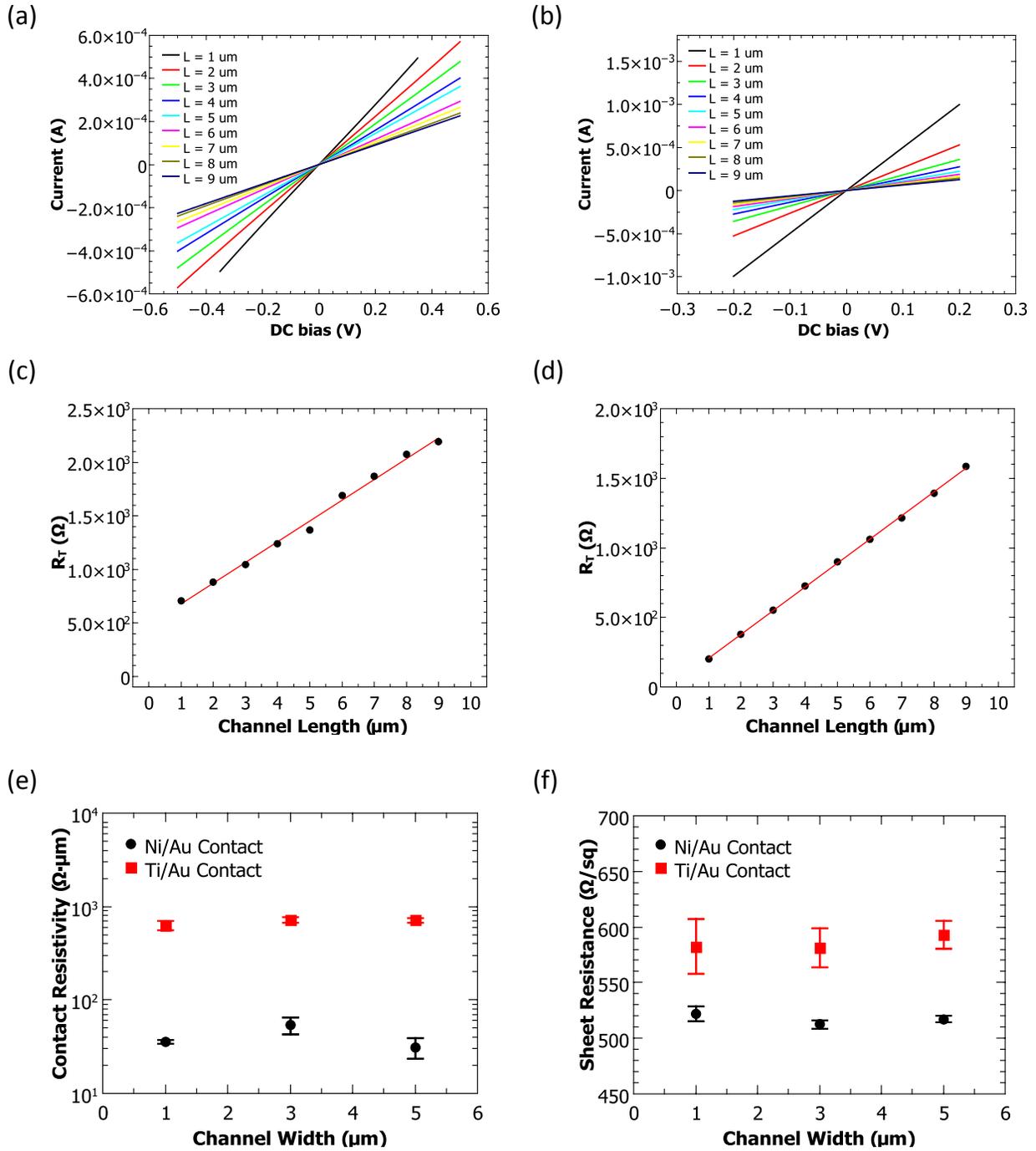

Figure 2. I – V plots of the TLM structure with various PtSe$_2$ channel length values (L) for (a) Ti/Au and (b) Ni/Au contacted devices, and the associated plots of the total resistance (R$_T$) vs.

PtSe$_2$ channel length for the (c) Ti/Au and (d) Ni/Au contacted devices. The initial Pt deposition thickness and the channel width of the devices are 5 nm and 3 μm, respectively. (e) Contact resistivity and (f) sheet resistance values extracted from the TLM measurements for the Ti/Au and Ni/Au contacted devices with different PtSe$_2$ channel width values.

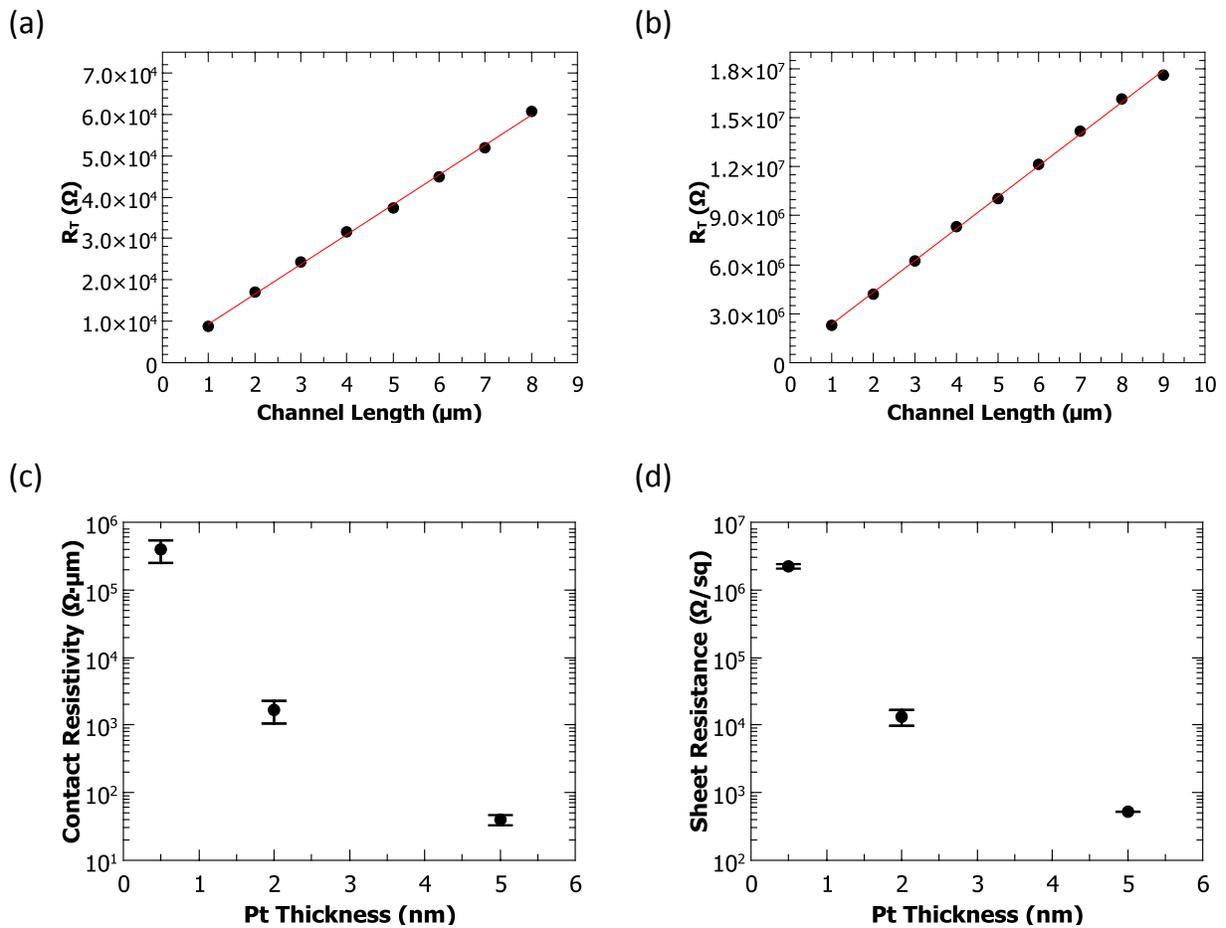

Figure 3. Plots of the total resistance (R$_T$) vs. PtSe$_2$ channel length for the Ni-contacted TLM structures with different PtSe$_2$ thicknesses, synthesized from the (a) 2 nm and (b) 0.5 nm thick Pt layers with 1 μm of channel width. Summary of (c) contact resistivity and (d) sheet resistance values extracted using TLM method for the different PtSe$_2$ thicknesses.

Ni-contacted TLM structures were then fabricated on PtSe$_2$ channels with different thicknesses, which were selenized from 0.5, 2 and 5 nm thick initial Pt layers. The Ni – PtSe$_2$ contact resistivity and sheet resistance were extracted by analyzing the TLM measurement data, R$_T$ vs. contact spacing, in Figure 3(a) and (b). The extracted values of the contact resistivity and sheet resistance are plotted against the film thickness in Figure 3(c) and (d). When the same Ni electrodes were deposited on the PtSe$_2$ channels with various thicknesses, devices with thinner PtSe$_2$ channels show much higher contact resistivity and larger sheet resistance. This can be attributed to the nature of layered PtSe$_2$, whereby the electronic character changes from semimetallic to semiconducting as the number of layers decreases.[27,30,31] The thinner PtSe$_2$ can be expected to be more semiconducting than the thicker, resulting in higher contact resistivity with metal electrodes as well as larger sheet resistance.

Charge transport measurements of PtSe$_2$ FETs were carried out at room temperature in ambient conditions to further investigate the electrical properties of PtSe$_2$ films with different thicknesses. PtSe$_2$ channels with a length of 5 µm and a width of 1 µm were probed via two separate metal source and drain electrodes, with gate biases applied to the silicon substrate in a back-gate configuration. The output characteristics of the FETs with PtSe$_2$ channels synthesized from 5, 2 and 0.5 nm thick initial Pt layers are plotted in Figure 4(a) – (c), respectively. For all devices, the drain – source current (I$_{ds}$) increases linearly with the applied dc drain – source voltage (V$_{ds}$), implying good ohmic contact of the Ni electrodes with the PtSe$_2$ channels. As expected, the device with a thicker PtSe$_2$ layer shows higher conductivity. The gating characteristics of the FETs were examined under a dc back-gate bias (V$_{gs}$) in the range of -80 to +80 volts (V). While the FETs with PtSe$_2$ channels synthesized from the 5 and 2 nm thick Pt layers show hardly any gate dependence, the PtSe$_2$ FET from the 0.5 nm thick initial Pt layer shows a clear gate dependence with p-type conduction. This is consistent with the previous results that contact formation with a high work function metal reduces the contact resistance at the metal-PtSe$_2$ junction. Also, this data supports the hypothesis that PtSe$_2$ becomes more semiconducting as it gets thinner. The field-effect mobility (µ) was extracted from the transfer characteristics of the PtSe$_2$ FETs selenized from the 0.5 nm thick Pt layer in Figure 4(d), using the expression µ = [L / (W × C$_{ox}$ × V$_{ds}$)] × [∂I$_{ds}$ / ∂V$_{gs}$], where L is the channel length, W is the channel width, C$_{ox}$ is the capacitance of the insulating layer between the gate and the channel. The mobility was estimated to be a maximum of 0.6 cm$^2$V$^{-1}$s$^{-1}$, which is lower than previously reported values (7 – 210 cm$^2$V$^{-1}$s$^{-1}$) obtained from high temperature CVD-grown single crystalline, or mechanically exfoliated PtSe$_2$.[38,39] However, considering the benefits of our growth process, namely the low synthesis temperature, scalability and ease of controlling layer thickness, this is quite striking. Such a relatively low mobility likely originates from the polycrystalline structure, with randomly distributed PtSe$_2$ grains, observed for TAC-grown PtSe$_2$.[32,33] Additional efforts to optimize the synthesis process, through the use of epitaxially deposited Pt or highly-crystalline substrates, and the realization of local top-gates with high-k dielectrics, are expected to improve the mobility. In order to assess the stability of PtSe$_2$ channels, we repeated I – V measurements on the same device (PtSe$_2$ channel with a length of 5 µm and a width of 1 µm) in ambient conditions with an interval of 20 days and monitored the variation of R$_T$ of the device for 40 days. As presented in Figure S4 of the Supplementary Information, the measured R$_T$ was 8.1 MΩ from the first measurements, 8.3 MΩ from the second measurements after 20 days, and 9.4 MΩ from the last measurements after 40 days. This indicates that the PtSe$_2$ channel is quite stable with a resistance increase of 15 % of the initial value after 40 days without any passivation or post-treatment.

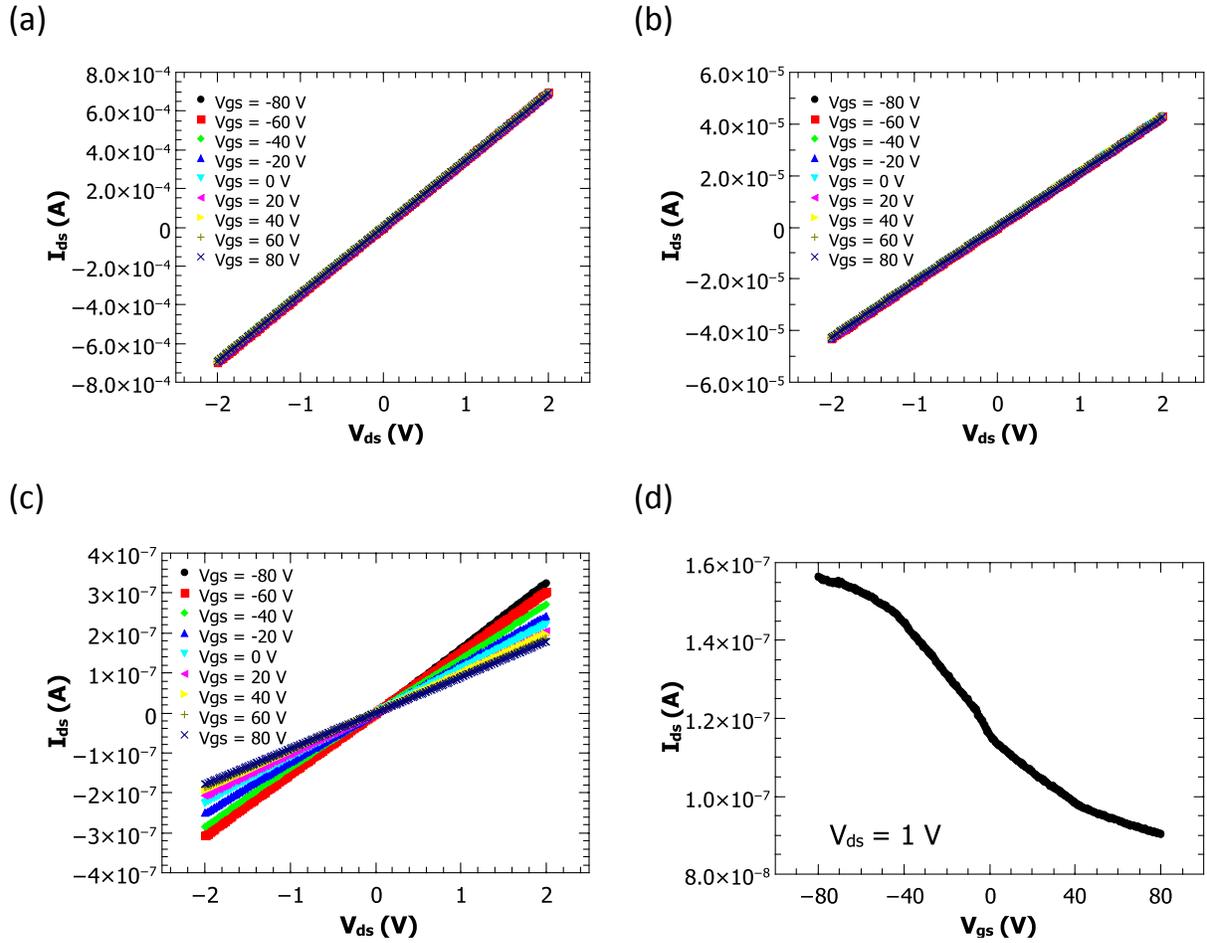

Figure 4. Output characteristics ($I_{ds}$ vs. $V_{ds}$) of PtSe$_2$ FET devices with PtSe$_2$ channels selenized from (a) 5 nm, (b) 2 nm and (c) 0.5 nm thick Pt layers under various back-side gate biases ($V_{gs}$) from -80 to 80 V. (d) Transfer characteristics ($I_{ds}$ vs. $V_{gs}$ at $V_{ds}$ = 1 V) of the PtSe$_2$ FET device with a PtSe$_2$ channel selenized from a 0.5 nm thick Pt layer.

Lastly, we investigated the effect of "edge contacts" on the contact resistance of PtSe$_2$ TLM structures. There have been continuous efforts to achieve low contact resistance in 2D-material based devices, including local plasma or ultraviolet/ozone treatment of the contact area[40,41] and molecular doping of the channel materials.[42] However, these methods require additional processes and can cause damage to the channel materials. Forming an "end-contacted" interface between metal and graphene was proposed as another way to reduce contact resistance,[43] wherein the end-contacted structure at the graphene – metal contacts facilitates chemical bonding between the graphene and the contact metal, enhancing the carrier injection at the contact area. Such an approach has previously been experimentally adopted for graphene-based devices, resulting in a significant improvement in the contact resistance.[44–46] Based on this principle, we expect that the edge-contacted structure shown here improves the carrier injection at the metal – PtSe$_2$ interface. Arrays of holes with a hole-size of 200 × 200 nm were patterned by EBL on the contact area of the channels, as presented in Figure 5(a) and (b). This leads to the formation of edge contacts between the PtSe$_2$ channel and metal electrodes and increases the total area of exposed channel edges at the contact region in the TLM structures. The PtSe$_2$ channels of all the devices were selenized from 0.5 nm thick Pt layers with a width of 2 μm, and three different types of hole-array patterns (side 1-line, side 2-line and center 2-line) were fabricated with Ni electrodes. The contact resistivity of the normal PtSe$_2$ TLM device without holes and the device with holes at the contact region were extracted and are compared in Figure 5(c). While there is no clear dependence of the contact resistivity on the number and location of hole-arrays at the contact region, distributed between 2 × 10$^{-5}$ and 6 × 10$^{-5}$ Ω·μm, all the devices with the hole-patterns at the contact region exhibit 50 – 70 % lower contact resistivity than the conventional device. In contrast, no significant difference in the sheet resistance was observed for all the PtSe$_2$ channel devices in Figure 5(d), indicating that the quality of the PtSe$_2$ channels are nearly constant for all devices.

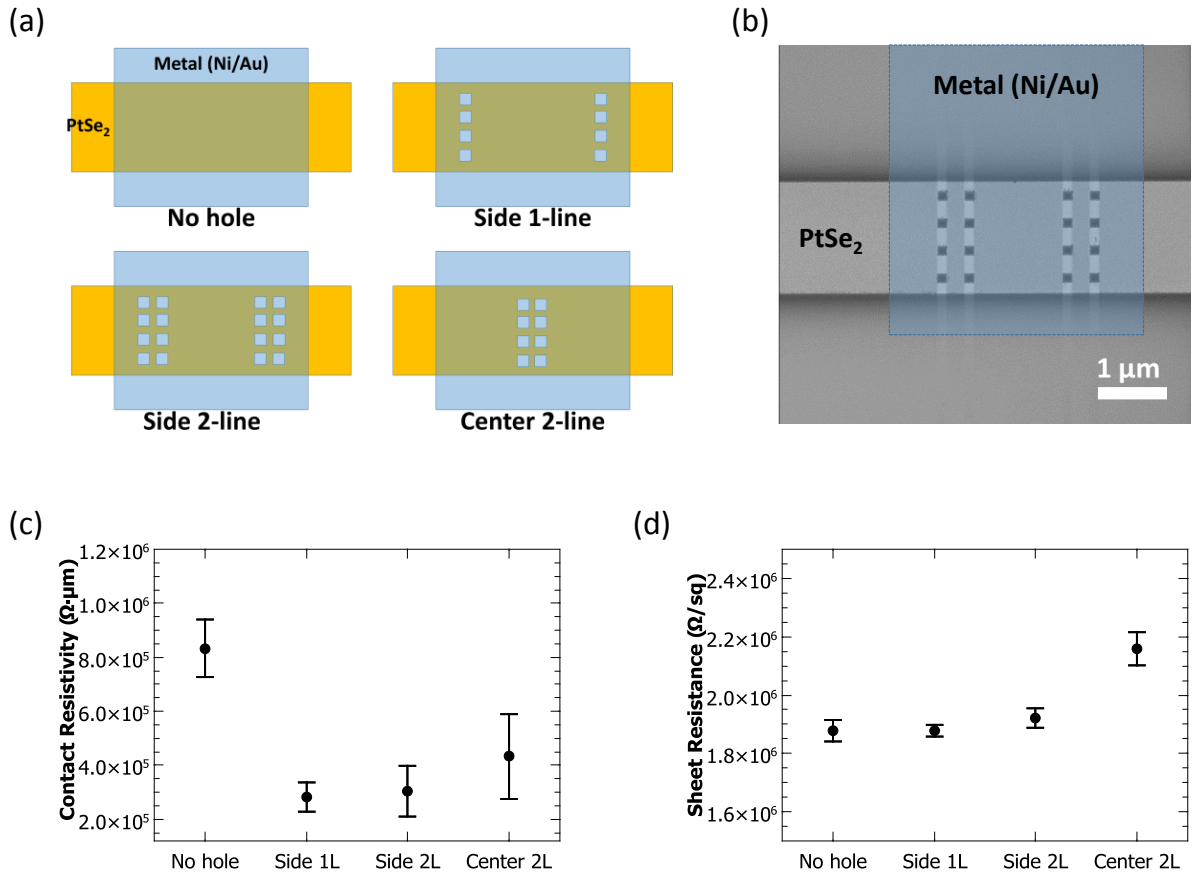

Figure 5. (a) Schematic diagram of the metal (Ni) – PtSe$_2$ contact area with/without holes on the PtSe$_2$ channel. (b) SEM image of the PtSe$_2$ channel with 2-line holes at the side of the metal contact area before metal electrode deposition. Extracted (c) contact resistivity and (d) sheet resistance values from the PtSe$_2$ TLM structures (0.5 nm of starting Pt thickness, 2 μm of channel width) with and without holes on the PtSe$_2$ channel at the metal-PtSe$_2$ contact area.

**Conclusions**

We have investigated the electrical contact properties of PtSe$_2$ channels by the TLM method. Initial Pt layers with thicknesses of 0.5, 2 and 5 nm were selenized at low temperature using a TAC method, realizing robust PtSe$_2$ layers with different thicknesses. When comparing the contact resistivity values of the PtSe$_2$ TLM devices with two different contact metals, Ti and Ni, it was found that Ni forms a better electrical contact to the PtSe$_2$ channel. We also observed that the thinner PtSe$_2$ films have a higher contact resistivity and larger sheet resistance from the TLM measurements, implying that thinner PtSe$_2$ films become more semiconducting. In addition, the charge transport characteristics of PtSe$_2$ FETs were investigated. Only the device derived from the thinnest PtSe$_2$ layer (0.5 nm) showed a clear gate dependence with p-type conduction, which confirms the transition of PtSe$_2$ from semimetal to semiconductor with decreasing thickness. Furthermore, the effect of edge-contacted structures on the contact resistance was examined. Arrays of holes were patterned in PtSe$_2$ to increase the "edge-contacted" area of the layered PtSe$_2$ film at the metal interface. We found that the edge-contacted structures reduce the contact resistivity, which we attribute to enhancement of the carrier injection at the contacts. Though more comprehensive studies should be carried out in the future to fully understand the fundamentals of the electrical contact properties, our findings provide a quick insight into the realization of high-performance nanoelectronic devices based on layered PtSe$_2$.

**Acknowledgements**

Support from an ERC grant (InteGraDe, 307311), the German Research Foundation (DFG, LE 2440/2-1), European Regional Funds (HEA2D), the French RENATECH network and Science Foundation Ireland (15/SIRG/3329 & 15/IA/3131) is gratefully acknowledged. V.P. would like to acknowledge Prof. Henri Happy for providing access to the cleanroom facilities and fruitful discussions.

# References

[1] Novoselov K S, Geim A K, Morozov S V, Jiang D, Zhang Y, Dubonos S V, Grigorieva I V and Firsov A A 2004 Electric Field Effect in Atomically Thin Carbon Films *Science* **306** 666–9

[2] Chhowalla M, Shin H S, Eda G, Li L-J, Loh K P and Zhang H 2013 The chemistry of two-dimensional layered transition metal dichalcogenide nanosheets *Nat. Chem.* **5** 263–75

[3] Fiori G, Bonaccorso F, Iannaccone G, Palacios T, Neumaier D, Seabaugh A, Banerjee S K and Colombo L 2014 Electronics based on two-dimensional materials *Nat. Nanotechnol.* **9** 768–79

[4] Rhoderick E H 1982 Metal-semiconductor contacts *IEE Proc. - Solid-State Electron Devices* **129** 1–14

[5] Castro Neto A H, Guinea F, Peres N M R, Novoselov K S and Geim A K 2009 The electronic properties of graphene *Rev. Mod. Phys.* **81** 109–62

[6] Robinson J A, LaBella M, Zhu M, Hollander M, Kasarda R, Hughes Z, Trumbull K, Cavalero R and Snyder D 2011 Contacting graphene *Appl. Phys. Lett.* **98** 053103

[7] Malec C E, Elkus B and Davidović D 2011 Vacuum-annealed Cu contacts for graphene electronics *Solid State Commun.* **151** 1791–3

[8] Zhong H, Zhang Z, Chen B, Xu H, Yu D, Huang L and Peng L 2015 Realization of low contact resistance close to theoretical limit in graphene transistors *Nano Res.* **8** 1669–79

[9] Gahoi A, Wagner S, Bablich A, Kataria S, Passi V and Lemme M C 2016 Contact resistance study of various metal electrodes with CVD graphene *Solid-State Electron.* **125** 234–9

[10] Radisavljevic B, Radenovic A, Brivio J, Giacometti V and Kis A 2011 Single-layer $MoS_2$ transistors *Nat. Nanotechnol.* **6** 147–50

[11] Ovchinnikov D, Allain A, Huang Y-S, Dumcenco D and Kis A 2014 Electrical Transport Properties of Single-Layer $WS_2$ *ACS Nano* **8** 8174–8181

[12] Lopez-Sanchez O, Lembke D, Kayci M, Radenovic A and Kis A 2013 Ultrasensitive photodetectors based on monolayer $MoS_2$ *Nat. Nanotechnol.* **8** 497–501

[13] Koppens F H L, Mueller T, Avouris P, Ferrari A C, Vitiello M S and Polini M 2014 Photodetectors based on graphene, other two-dimensional materials and hybrid systems *Nat. Nanotechnol.* **9** 780–93

[14]  Yim C, O'Brien M, McEvoy N, Riazimehr S, Schafer-Eberwein H, Bablich A, Pawar R, Iannaccone G, Downing C, Fiori G, Lemme M C and Duesberg G S 2014 Heterojunction Hybrid Devices from Vapor Phase Grown MoS$_2$ *Sci. Rep.* **4** 5458

[15]  Lee K, Gatensby R, McEvoy N, Hallam T and Duesberg G S 2013 High Performance Sensors Based on Molybdenum Disulfide Thin Films *Adv. Mater.* **25** 6699–702

[16]  Perkins F K, Friedman A L, Cobas E, Campbell P M, Jernigan G G and Jonker B T 2013 Chemical Vapor Sensing with Monolayer MoS$_2$ *Nano Lett.* **13** 668–73

[17]  Das S, Chen H-Y, Penumatcha A V and Appenzeller J 2013 High Performance Multilayer MoS$_2$ Transistors with Scandium Contacts *Nano Lett.* **13** 100–5

[18]  Allain A, Kang J, Banerjee K and Kis A 2015 Electrical contacts to two-dimensional semiconductors *Nat. Mater.* **14** 1195–205

[19]  Chuang H-J, Chamlagain B, Koehler M, Perera M M, Yan J, Mandrus D, Tománek D and Zhou Z 2016 Low-Resistance 2D/2D Ohmic Contacts: A Universal Approach to High-Performance WSe$_2$, MoS$_2$, and MoSe$_2$ Transistors *Nano Lett.* **16** 1896–902

[20]  Yoon Y, Ganapathi K and Salahuddin S 2011 How Good Can Monolayer MoS$_2$ Transistors Be? *Nano Lett.* **11** 3768–73

[21]  Kwon H-J, Jang J, Kim S, Subramanian V and Grigoropoulos C P 2014 Electrical characteristics of multilayer MoS$_2$ transistors at real operating temperatures with different ambient conditions *Appl. Phys. Lett.* **105** 152105

[22]  Chen B, Sahin H, Suslu A, Ding L, Bertoni M I, Peeters F M and Tongay S 2015 Environmental Changes in MoTe$_2$ Excitonic Dynamics by Defects-Activated Molecular Interaction *ACS Nano* **9** 5326–32

[23]  Shimazu Y, Tashiro M, Sonobe S and Takahashi M 2016 Environmental Effects on Hysteresis of Transfer Characteristics in Molybdenum Disulfide Field-Effect Transistors *Sci. Rep.* **6** 30084

[24]  Lee Y-H, Zhang X-Q, Zhang W, Chang M-T, Lin C-T, Chang K-D, Yu Y-C, Wang J T-W, Chang C-S, Li L-J and Lin T-W 2012 Synthesis of Large-Area MoS$_2$ Atomic Layers with Chemical Vapor Deposition *Adv. Mater.* **24** 2320–5

[25]  Elías A L, Perea-López N, Castro-Beltrán A, Berkdemir A, Lv R, Feng S, Long A D, Hayashi T, Kim Y A, Endo M, Gutiérrez H R, Pradhan N R, Balicas L, Mallouk T E, López-Urías F, Terrones H and Terrones M 2013 Controlled Synthesis and Transfer of Large-Area WS$_2$ Sheets: From Single Layer to Few Layers *ACS Nano* **7** 5235–5242

[26]  Gatensby R, McEvoy N, Lee K, Hallam T, Berner N C, Rezvani E, Winters S, O'Brien M and Duesberg G S 2014 Controlled synthesis of transition metal dichalcogenide thin films for electronic applications *Appl. Surf. Sci.* **297** 139–46

[27] Miró P, Ghorbani-Asl M and Heine T 2014 Two Dimensional Materials Beyond MoS$_2$: Noble-Transition-Metal Dichalcogenides *Angew. Chem. Int. Ed.* **53** 3015–8

[28] Huang Z, Zhang W and Zhang W 2016 Computational Search for Two-Dimensional MX$_2$ Semiconductors with Possible High Electron Mobility at Room Temperature *Materials* **9** 716

[29] Guo G Y and Liang W Y 1986 The electronic structures of platinum dichalcogenides: PtS$_2$, PtSe$_2$ and PtTe$_2$ *J. Phys. C Solid State Phys.* **19** 995

[30] Zhuang H L and Hennig R G 2013 Computational Search for Single-Layer Transition-Metal Dichalcogenide Photocatalysts *J. Phys. Chem. C* **117** 20440–5

[31] Wang Y, Li L, Yao W, Song S, Sun J, Pan J, Ren X, Li C, Okunishi E, Wang Y-Q, Wang E, Shao Y, Zhang Y, Yang H, Schwier E F, Iwasawa H, Shimada K, Taniguchi M, Cheng Z, Zhou S, Du S, Pennycook S J, Pantelides S T and Gao H-J 2015 Monolayer PtSe$_2$, a New Semiconducting Transition-Metal-Dichalcogenide, Epitaxially Grown by Direct Selenization of Pt *Nano Lett.* **15** 4013–8

[32] O'Brien M, McEvoy N, Motta C, Zheng J-Y, Berner N C, Kotakoski J, Kenan Elibol, Pennycook T J, Meyer J C, Yim C, Abid M, Hallam T, Donegan J F, Sanvito S and Duesberg G S 2016 Raman characterization of platinum diselenide thin films *2D Mater.* **3** 021004

[33] Yim C, Lee K, McEvoy N, O'Brien M, Riazimehr S, Berner N C, Cullen C P, Kotakoski J, Meyer J C, Lemme M C and Duesberg G S 2016 High-Performance Hybrid Electronic Devices from Layered PtSe$_2$ Films Grown at Low Temperature *ACS Nano* **10** 9550–8

[34] Sedky S, Witvrouw A, Bender H and Baert K 2001 Experimental determination of the maximum post-process annealing temperature for standard CMOS wafers *IEEE Trans. Electron Devices* **48** 377–85

[35] Takeuchi H, Wung A, Sun X, Howe R T and King T-J 2005 Thermal budget limits of quarter-micrometer foundry CMOS for post-processing MEMS devices *IEEE Trans. Electron Devices* **52** 2081–6

[36] Lee Y H D and Lipson M 2013 Back-End Deposited Silicon Photonics for Monolithic Integration on CMOS *IEEE J. Sel. Top. Quantum Electron.* **19** 8200207

[37] Reeves G K and Harrison H B 1982 Obtaining the specific contact resistance from transmission line model measurements *IEEE Electron Device Lett.* **3** 111–3

[38] Wang Z, Li Q, Besenbacher F and Dong M 2016 Facile Synthesis of Single Crystal PtSe$_2$ Nanosheets for Nanoscale Electronics *Adv. Mater.* **28** 10224–9

[39]  Zhao Y, Qiao J, Yu Z, Yu P, Xu K, Lau S P, Zhou W, Liu Z, Wang X, Ji W and Chai Y 2017 High-Electron-Mobility and Air-Stable 2D Layered PtSe$_2$ FETs *Adv. Mater.* **29** 1604230

[40]  Choi M S, Lee S H and Yoo W J 2011 Plasma treatments to improve metal contacts in graphene field effect transistor *J. Appl. Phys.* **110** 073305

[41]  Li W, Liang Y, Yu D, Peng L, Pernstich K P, Shen T, Hight Walker A R, Cheng G, Hacker C A, Richter C A, Li Q, Gundlach D J and Liang X 2013 Ultraviolet/ozone treatment to reduce metal-graphene contact resistance *Appl. Phys. Lett.* **102** 183110

[42]  Yang L, Majumdar K, Liu H, Du Y, Wu H, Hatzistergos M, Hung P Y, Tieckelmann R, Tsai W, Hobbs C and Ye P D 2014 Chloride Molecular Doping Technique on 2D Materials: WS$_2$ and MoS$_2$ *Nano Lett.* **14** 6275–80

[43]  Matsuda Y, Deng W-Q and Goddard W A 2010 Contact Resistance for "End-Contacted" Metal−Graphene and Metal−Nanotube Interfaces from Quantum Mechanics *J. Phys. Chem. C* **114** 17845–50

[44]  Smith J T, Franklin A D, Farmer D B and Dimitrakopoulos C D 2013 Reducing Contact Resistance in Graphene Devices through Contact Area Patterning *ACS Nano* **7** 3661–7

[45]  Leong W S, Gong H and Thong J T L 2014 Low-Contact-Resistance Graphene Devices with Nickel-Etched-Graphene Contacts *ACS Nano* **8** 994–1001

[46]  Passi V, Gahoi A, Ruhkopf J, Kataria S, Vaurette F, Pallecchi E, Happy H and Lemme M C 2016 Contact Resistance Study of "Edge-Contacted" Metal-Graphene Interfaces *46th European Solid-State Device Research Conference (ESSDERC), 2016* 46th European Solid-State Device Research Conference (ESSDERC), 2016 (IEEE) pp 236–9

**Supplementary Information**

**SEM characterization of PtSe$_2$ layers**

Scanning electron microscopy (SEM) was used to image PtSe$_2$ films grown from Pt films with starting thicknesses of 0.5, 2 and 5 nm. This was performed using a Zeiss Ultra Plus FE SEM operated at 2 kV using an in-lens detector. The acquired images, in which the polycrystalline nature of the films grown is evident, are shown in Fig S1.

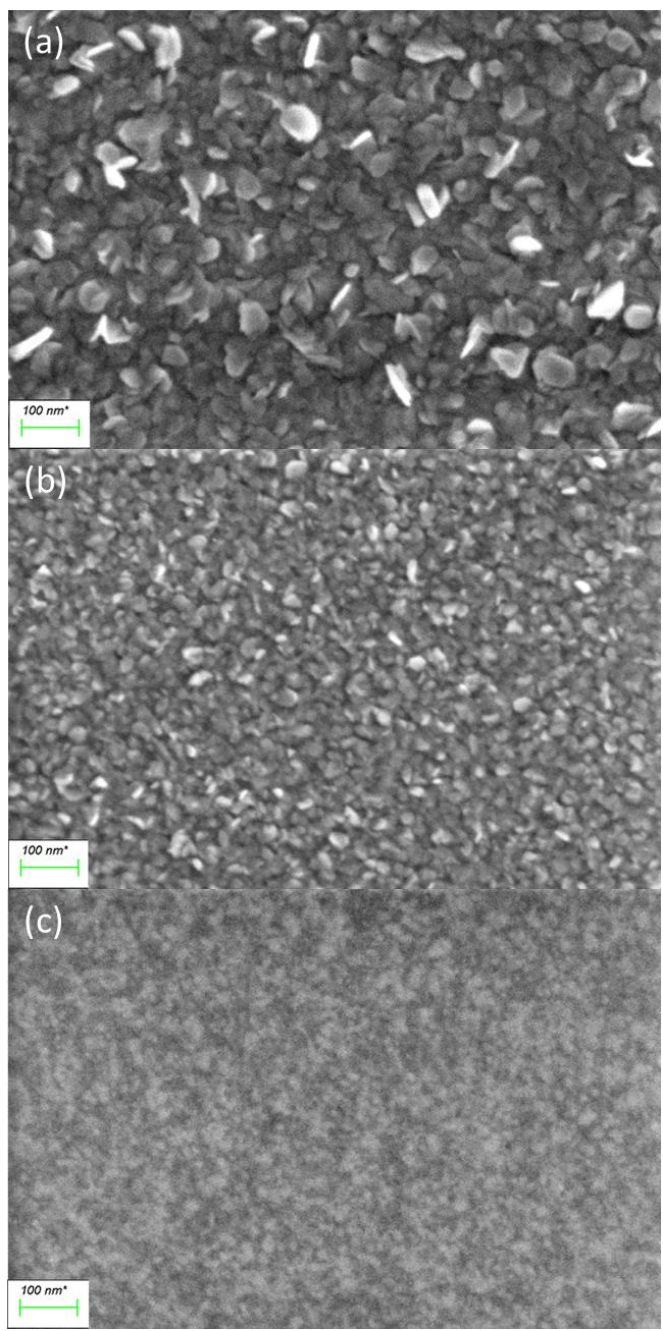

Figure S1. Scanning electron microscopy (SEM) images of PtSe$_2$ films grown from Pt of starting thickness (a) 5 nm, (b) 2 nm and (c) 0.5 nm.

**AFM characterization of Pt and PtSe$_2$ layers**

Atomic force microscopy (AFM) imaging was conducted on Pt samples with starting thicknesses of 0.5, 2 and 5 nm, as measured by the sputtering tool's quartz crystal microbalance (QCM), both pre- and post selenization, using a NT-MDT Solver PRO operating in tapping mode. Edge regions, produced using a shadow mask during deposition of the Pt, were chosen so that line profiles could be used to determine the height of the films. Images of these regions and the inset associated line profiles are shown in Fig S2. (a-f). A plot of Pt thickness versus resultant PtSe$_2$ thickness, both determined from AFM line profiles, is shown in Fig S3 (a) and indicates that the films expand by a factor of ~4 upon selenization. A table with thickness values for Pt films measured by the quartz crystal microbalance (QCM) of the sputtering tool, thickness values for the Pt films measured by AFM and thickness values for the resultant PtSe$_2$ films measured by AFM is shown in Fig S3. (b).

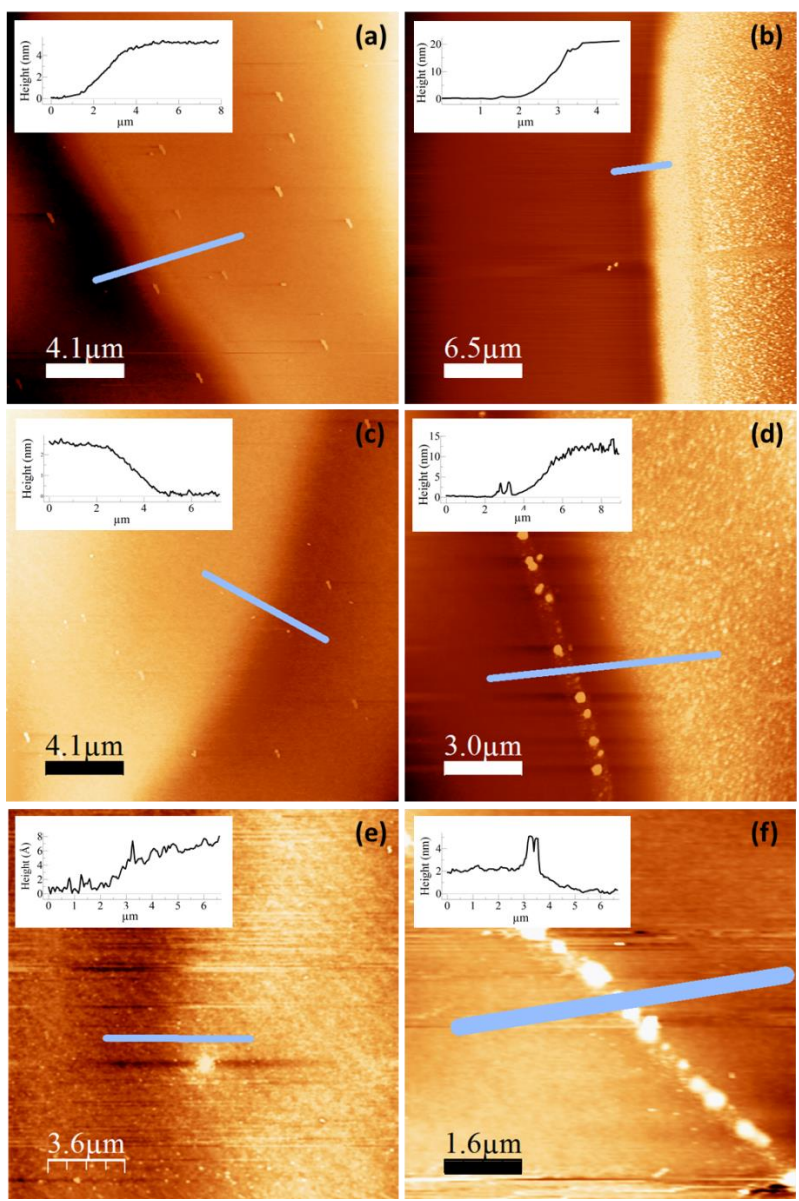

Figure S2. (a, c, e) AFM images of a Pt films with starting thicknesses of 5, 2 and 0.5 nm, respectively, as measured by the QCM. (b, d, f) Corresponding images of the same films post selenization. Inset: extracted line profiles.

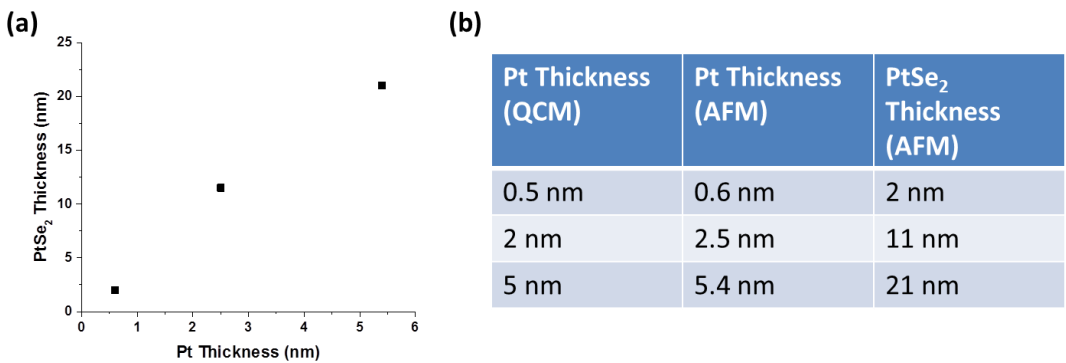

Figure S3 (a) Pt thickness plotted versus post-selenization PtSe$_2$ thickness (b) Table of thickness values measured.

**Monitoring stability of the PtSe$_2$ channel.**

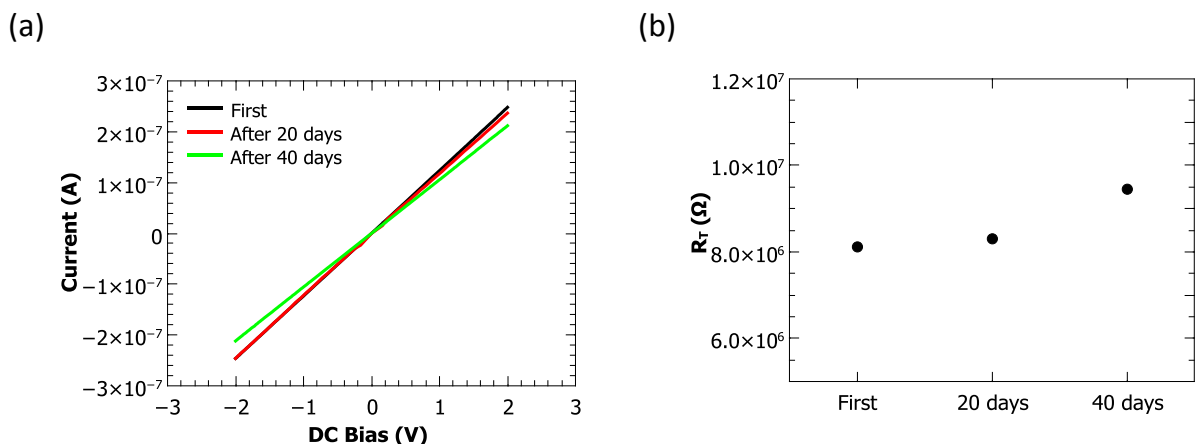

Figure S4. (a) I – V plots and (b) total resistance values of the two-electrode PtSe$_2$ channel, repeatedly measured in ambient conditions with an interval of 20 days for 40 days.